\documentclass[lettersize,journal]{IEEEtran}
\IEEEoverridecommandlockouts
\usepackage{cite}
\usepackage{amsmath,amssymb,amsfonts}
\usepackage{algorithmic}
\usepackage{graphicx}
\usepackage{textcomp}
\usepackage{glossaries}
\usepackage{siunitx}
\usepackage{changepage}
\usepackage{caption}
\usepackage{subcaption}

\usepackage{tikz}
\usetikzlibrary{arrows.meta, positioning,patterns}
\usetikzlibrary{patterns.meta}

\usepackage{circuitikz, spath3}
\usetikzlibrary{
    spath3,
    intersections,
    shapes.geometric,
}

\usetikzlibrary{shapes,arrows,positioning,calc}

\tikzset{
block/.style = {draw, fill=white, rectangle, minimum height=3em, minimum width=3em},
bigblock/.style = {draw, fill=white, rectangle, minimum height=25em, minimum width=6em},
bigblock2/.style = {draw, fill=white, rectangle, minimum height=15em, minimum width=6em},,
tmp/.style  = {coordinate}, 
sum/.style= {draw, fill=white, circle, node distance=2cm},
input/.style = {coordinate},
output/.style= {coordinate},
pinstyle/.style = {pin edge={to-,thin,black},
mult/.style={circle, draw, minimum size=6mm, inner sep=0pt,
node contents={\huge$\times$}},
}
}

\newacronym{FAS}{FAS}{fluid antenna system}
\newacronym{SNR}{SNR}{signal to noise ratio}
\newacronym{GP}{GP}{Gaussian processes}
\newacronym{OP}{OP}{Outage Probability}
\newacronym{BLUE}{BLUE}{best linear unbiased estimate}
\newacronym{NMSE}{NMSE}{normalized mean squared error}
\newacronym{MMSE}{MMSE}{minimum mean square error}
\newacronym{i.i.d.}{i.i.d}{independent and identically distributed}
\newacronym{CSI}{CSI}{channel state information}
\newacronym{CSIT}{CSIT}{transimitter-side channel state information}
\newacronym{RF}{RF}{radio frequency}
\newacronym{MT}{MT}{mobile terminal}
\newacronym{BS}{BS}{base station}
\newacronym{LLR}{LLR}{log likelihood ratio}
\newacronym{QPSK}{QPSK}{quadrature phase shift keying}
\newacronym{PSK}{PSK}{phase shift keying}
\newacronym{BER}{BER}{bit error rate}
\newacronym{OFDM}{OFDM}{orthogonal frequency-division multiplexing}
\newacronym{MIMO}{MIMO}{multiple input multiple output}
\newacronym{NL}{NL}{nonlinear}
\newacronym{IQ}{IQ}{in-phase quadrature}
\newacronym{SDR}{SDR}{signal-to-distortion ratio}
\newacronym{SINR}{SINR}{signal-to-interference noise ratio}
\newacronym{IMP}{IMP}{intermodulation product}
\newacronym{SVD}{SVD}{singular value decomposition}
\newacronym{s-FAMA}{s-FAMA}{slow-FAMA}
\newacronym{DDIR}{DDIR}{decision-directed iterative receiver}
\newacronym{GAB}{GAB}{genie-aided bound}
\newacronym{ML}{ML}{maximum-likelihood}
\newacronym{QAM}{QAM}{quadrature amplitude modulation}
\newacronym{GAMP}{GAMP}{generalized approximate message passing}
\newacronym{PAPR}{PAPR}{peak-to-average power ratio}
\newacronym{PA}{PA}{power amplifier}
\newacronym{FFT}{FFT}{fast Fourier transform}
\newacronym{DFT}{DFT}{discrete Fourier transform}
\newacronym{CLT}{CLT}{central limit theorem}
\newacronym{IM}{IM}{index modulation}
\newacronym{FASCC}{FASCC}{fluid antenna secure channel-driven constellation}
\newacronym{DU-FASCC}{DU-FASCC}{deliberate uncertaintly-FASCC}
\newacronym{DC-FASCC}{DC-FASCC}{destructive combining-FASCC}
\newacronym{PLS}{PLS}{physical layer security}
\newacronym{FAMA}{FAMA}{fluid antenna multiple access}
\newacronym{5G}{5G}{fifth generation}
\newacronym{6G}{6G}{sixth generation}
\newacronym{NOMA}{NOMA}{non-orthogonal multiple access}

\usepackage[dvipsnames]{xcolor}
\def\BibTeX{{\rm B\kern-.05em{\sc i\kern-.025em b}\kern-.08em
    T\kern-.1667em\lower.7ex\hbox{E}\kern-.125emX}}

\begin{document}
\bstctlcite{BSTcontrol}

\title{Fluid Antenna Secure Channel-driven Constellation (FASCC): a Physical Layer Security approach
}

\author{\IEEEauthorblockN{Daniel Dinis , Risto Wichman}\\
\thanks{D. Dinis and R. Wichman are with the 
School of Electrical Engineering, Aalto University; (emails: \{daniel.1.dinis, risto.wichman\}@aalto.fi)}


\thanks{
 The work was supported by Research Council of Finland Grant 367569.}
}

\maketitle

\begin{abstract}
\Gls{FAS} have recently emerged as a promising technology for future wireless networks due to their ability to access a large number of channel realizations using a single radio-frequency chain. This capability creates new opportunities for physical-layer security beyond conventional beamforming and artificial-noise approaches. In this paper, we propose a novel \gls{FASCC} framework for downlink communications, in which information is conveyed through channel-domain constellation points generated by FAS port selection and phase adaptation. 
Two complementary \gls{FASCC} schemes are developed, depending on the availability of eavesdropper's \gls{CSI}. Numerical results demonstrate that the proposed \gls{FASCC} framework achieves significant secrecy gains in terms of \gls{BER} and mutual-information secrecy performance, even when Eve is significantly closer to the \gls{BS} than Bob.
\end{abstract}


\section{Introduction}

The broadcast nature of wireless communications makes transmitted information inherently vulnerable to eavesdropping. As wireless networks continue to evolve toward beyond-\gls{5G} and \gls{6G} systems supporting massive connectivity, immersive applications, and mission-critical services, ensuring secure transmission has become increasingly important\cite{6G_study2020}. While conventional cryptographic techniques provide a fundamental layer of protection, they typically rely on computational assumptions and key-management procedures that may become challenging in large-scale and resource-constrained networks. Consequently, \gls{PLS} has emerged as a complementary approach that exploits the randomness of wireless propagation channels to enhance communication confidentiality\cite{Liu2017_PLS_survey}.

\Gls{MIMO} technology plays a central role in modern wireless systems and has become one of the primary enablers of \gls{PLS}\cite{Goel2008_MIMO_PLS}. By exploiting spatial degrees of freedom, multi-antenna systems can improve the quality of legitimate links while limiting information leakage to potential eavesdroppers\cite{Son2024_PLS_MIMO}. However, the deployment of large antenna arrays is often accompanied by increased hardware cost, power consumption, and signal-processing complexity due to the need for multiple radio-frequency chains\cite{Bjornson2017}.

To address these challenges, \gls{FAS} have recently emerged as a promising antenna architecture for future wireless networks \cite{Wong2021,WongTutorial2025}. Unlike conventional antenna arrays, \gls{FAS} enable the dynamic repositioning of a radiating element across a set of densely distributed ports within a finite aperture. Early implementations include mechanically actuated systems and liquid antennas \cite{Bakar2021,Zhu2024mov}, while more recent advances, such as reconfigurable pixel structures and meta-fluid antennas, have been proposed to achieve faster and more flexible port switching \cite{Zhang2025_pixel,Liu2025_MetaFluid}. By providing access to a large number of spatially distinct channel realizations using a single radio-frequency chain, \gls{FAS} introduce new degrees of freedom that can substantially enhance wireless system performance \cite{WongTutorial2025}. Consequently, \gls{FAS} have attracted significant research interest in areas such as channel estimation \cite{Xu2025sparse,Dinis2026_time}, beamforming \cite{Ma2024_Beamforming}, and multi-user transmission schemes including \gls{FAMA} \cite{Wong2022_1FAMA,Dinis2026_sFAMAGP} and \gls{NOMA} \cite{ghadi24noma}. Furthermore, the large number of channel realizations available through port selection creates new opportunities for physical-layer security.

Several recent works have investigated the use of \gls{FAS} for secure communications. Existing studies have analyzed secrecy metrics such as average secrecy capacity and secrecy outage probability\cite{Ghadi2024,Wu2026_PLS_FAS}, while others have developed secure transmission techniques inspired by conventional MIMO-based \gls{PLS} approaches, including beamforming\cite{Hu2024_PLS_FAS}, artificial-noise generation\cite{Tang2023_PLS_FAS}, and port-selection strategies. These methods exploit the large number of channel realizations available through \gls{FAS} to enhance the received signal quality at Bob while suppressing or disrupting Eve's reception. Although promising, most of these approaches rely on the availability of Eve's \gls{CSI} to optimize port selection, beamforming weights, or power allocation. Since eavesdroppers are typically passive and do not cooperate with the legitimate system, acquiring such information is generally impractical\cite{Liu2017_PLS_survey}. Consequently, the design of FAS-based secure transmission schemes that can operate effectively without Eve's \gls{CSI} remains an important open problem.

More recently, \gls{FAS} have been combined with \gls{IM}, where part of the transmitted information is conveyed through the selected antenna port\cite{Chen2024_PLS_FAS_IM}. Owing to the large number of available \gls{FAS} ports, these approaches can achieve attractive spectral-efficiency gains while naturally introducing a degree of transmission uncertainty. However, the security benefits reported by existing \gls{FAS}-\gls{IM} schemes often rely on the assumption that the mapping between antenna ports and information bits is unknown to Eve. In practice, this assumption is difficult to justify against sophisticated adversaries capable of estimating or inferring the underlying transmission structure.



\subsection{Major Contributions}

Motivated by these observations, this paper proposes a novel \gls{FASCC} framework for downlink communications, in which information is conveyed through channel-domain constellation points generated by \gls{FAS} port selection and phase adaptation. Unlike conventional \gls{FAS}-based \gls{IM} schemes, the proposed framework does not rely on the secrecy of the port-to-bit mapping. Instead, security is embedded directly into the geometry of the received constellations observed by Bob and Eve. 

To realize this concept, two complementary \gls{FASCC} transmission strategies are developed. The first operates using only Bob's \gls{CSI} and exploits multiple transmission realizations to increase Eve's uncertainty, whereas the second additionally leverages Eve's \gls{CSI} to covert communication. 
Furthermore, an artificial transmit perturbation mechanism is introduced to further enhance security. The main contributions of this paper are summarized as follows:

\begin{itemize}

\item We propose a novel \gls{FASCC} framework that conveys information through channel-domain constellation points generated by \gls{FAS} port selection and phase adaptation. Unlike existing \gls{FAS}-based index modulation schemes, the proposed approach embeds security directly into the received constellation geometry and does not rely on secrecy of the port-to-bit mapping.

\item We develop a \gls{DU-FASCC} scheme that operates using only Bob's \gls{CSI}. By associating multiple transmission realizations with each message and randomly selecting among them during transmission, \gls{DU-FASCC} preserves a wide constellation at Bob while increasing the ambiguity of Eve's observations.

\item We develop a \gls{DC-FASCC} scheme that exploits both Bob's and Eve's \gls{CSI} to jointly optimize the selection of multiple active ports and their associated phase weights. By coherently combining the transmitted signals at Bob while inducing destructive combining at Eve, the proposed design maximizes constellation distinguishability at the legitimate receiver while driving the received signal at Eve toward the noise floor, significantly reducing the eavesdropper's ability to detect the presence of an ongoing transmission.


\item We provide extensive numerical results demonstrating substantial secrecy gains achieved by the proposed \gls{FASCC} framework in terms of \gls{BER} and mutual-information secrecy performance for different \gls{FAS} architectures.

\end{itemize}

\subsection{Paper Organization}

The remainder of this paper is organized as follows. Section \ref{sec:system} presents the system model, including the channel characterization and receiver detection framework. Section \ref{sec:architecture} introduces the proposed \gls{FASCC} transmission schemes and discusses the associated design considerations and performance metrics. Section \ref{sec:Performance} presents the numerical results and provides a detailed performance analysis. Finally, conclusions are drawn in Section \ref{sec:conclusions}.


\section{System Model}
\label{sec:system}

\begin{figure}[!h]
\centering
\input{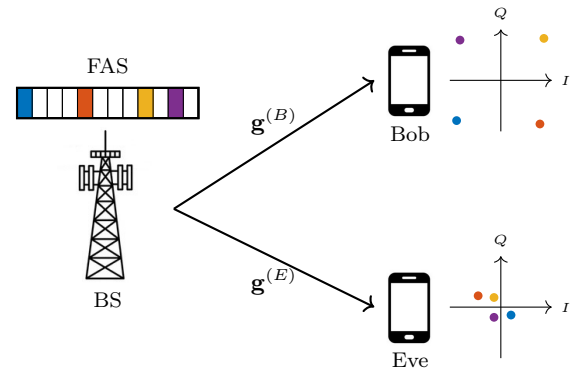}
\caption{Illustration of the proposed FAS-based secure communication system.}
\label{fig:comm_system}
\end{figure}

We consider a wireless downlink secure communication system consisting of a \gls{BS}, a legitimate receiver (Bob), and an eavesdropper (Eve). Both Bob and Eve are equipped with a single receive antenna. The \gls{BS} employs a \gls{FAS} capable of dynamically selecting one among $N$ ports distributed along a linear aperture of length $W\lambda$, where $\lambda$ denotes the carrier wavelength. The considered system model is illustrated in Fig.~\ref{fig:comm_system}, highlighting how the FAS-equipped BS creates distinct channel-domain constellations at Bob and Eve.

At the beginning of each coherence interval $T_s$, pilot signals are transmitted, enabling the estimation of the channel coefficients associated with each \gls{FAS} port\footnote{Several techniques have already been developed for efficient \gls{FAS} channel estimation\cite{Xu2025sparse,Dinis2026_time}.}. Since the pilots are broadcasted, Eve is also assumed to acquire perfect knowledge of her own channel coefficients corresponding to all \gls{FAS} ports. Based on the estimated channels, the \gls{BS} and Bob agree on the selected ports and bit-to-constellation mapping. The \gls{BS} subsequently activates the selected ports during data transmission.
Eve may also intercept the control signaling conveying the selected ports and bit-to-constellation mapping. However, since this mapping is designed based on Bob's channel realization, knowledge of the mapping alone does not enable Eve to reliably reproduce the intended constellation and, consequently, does not guarantee successful data decoding.


\subsection{Signal Model}
\label{sec:model}

The information bits are mapped into a message index
$m \in {1,\ldots,L}$,
where $L=2^b$ and $b$ denotes the number of transmitted bits.
Unlike conventional modulation schemes, where information is conveyed through the transmitted symbols, the proposed framework encodes the message index through the channel realization induced by the selected \gls{FAS} ports. Specifically, the transmitter exploits the channel coefficients across the $N$ available ports to construct a set of distinguishable channel-domain constellation points, each corresponding to a different message index. The specific construction of these constellation points depends on the adopted design algorithm and will be detailed in Section~\ref{sec:architecture}.

For a given message $m$, the selected port configuration $k$ and transmitted signal $s_{m,k}$ generate a message signal $x_m^{(u)}$ at receiver $u$, where $u\in\{B,E\}$ denotes Bob and Eve, respectively. Consequently, the received signal can be expressed as
\begin{equation}\label{eq:y_rx}
y^{(u)} = x_m^{(u)} + n^{(u)},
\end{equation}
where $n^{(u)} \sim \mathcal{CN}(0,\sigma_n^2)$ denotes additive white Gaussian noise.

Assuming perfect \gls{CSI}, the receiver performs maximum-likelihood detection according to
\begin{equation}
\hat{m}^{(u)} = \arg\min_{m} \left| y^{(u)} - x_m^{(u)} \right|^2.
\label{eq:ml_generic}
\end{equation}

The objective of the proposed designs is to maximize 
the Euclidian distance between $x_m^{(u)}$
at Bob while reducing their observability at Eve.

\subsection{FAS Channel Model}
\label{subsec:channel}

A block-fading channel is assumed, where the channel remains constant within each coherence interval. Each channel coefficient between an \gls{FAS} port $k$ and a receiver is modeled as a circularly symmetric complex Gaussian random variable,
$
g_{k}^{(u)} \sim \mathcal{CN}(0,\sigma^2),
$
with $\mathbb{E}[|g_{k}^{(u)}|^2]=\sigma^2=1$. Consequently, the channel magnitude $|g_{k}^{(u)}|$ follows a Rayleigh distribution for both Bob and Eve.

Due to the closely spaced ports in a \gls{FAS}, the channel coefficients across different ports are spatially correlated. Under the standard assumptions of narrowband flat fading and isotropic scattering, this spatial correlation can be described by the classical Jakes model~\cite{Jakes1994}. Assuming a linear fluid antenna with uniformly spaced ports, the spatial covariance between the channel coefficients associated with ports $k$ and $\ell$ is given by
\begin{align}
    (\mathbf{K})_{k,\ell} &= \mathbb{E}\left[g^{(u)}_k {g^{(u)}_\ell}^* \right] \nonumber = \sigma^2 J_0\left(2\pi \Delta d_{k,\ell} \right) \nonumber \\
    &= \sigma^2 J_0\left(\frac{2\pi (k - \ell)}{N-1} W \right),
    \label{eq:cova}
\end{align}
where $J_0(\cdot)$ is the zeroth-order Bessel function of the first kind and$\Delta d_{k,\ell}$ denotes the physical separation between ports $k$ and $\ell$.

The resulting covariance matrix $\mathbf{K}\in\mathbb{C}^{N\times N}$ characterizes the spatial correlation among the $N$ \gls{FAS} ports and is used to generate the channel vectors of Bob, $\mathbf{g}^{(B)}$, and Eve, $\mathbf{g}^{(E)}$.


\section{Fluid Antenna Secure Channel Modulation (FASCC)}
\label{sec:architecture}

As discussed in Section~\ref{sec:model}, each message index $m$ is associated with a message signal $x_m^{(u)}$, which acts as a channel-domain constellation point at receiver $u$. The performance of the proposed framework therefore depends on the construction of the set

\begin{equation}
\mathcal{X}^{(u)} = \left\{ x_1^{(u)},x_2^{(u)},\ldots,x_L^{(u)} \right\}.
\end{equation}

To this end, we propose the \gls{FASCC} framework, which exploits the spatial diversity provided by the $N$ \gls{FAS} ports to design the constellation points $\mathcal{X}^{(u)}$. By appropriately selecting the transmit ports, the proposed framework aims to maximize the distinguishability of the constellation points observed by Bob while simultaneously reducing the information that can be extracted by Eve. Depending on the available channel knowledge, this objective can be achieved through different constellation-design strategies.

In this work, two \gls{FASCC} schemes are proposed. The first scheme, referred to as \gls{DU-FASCC}, exploits the large number of available \gls{FAS} ports to generate multiple candidate channel realizations for each message. These additional degrees of freedom are then used to optimize the resulting constellation geometry at Bob without requiring knowledge of Eve's channel.

The second scheme, denoted as \gls{DC-FASCC}, assumes the availability of Eve's \gls{CSI}. In this case, the selected \gls{FAS} ports and transmit weights are jointly designed to achieve constructive combining at Bob and destructive combining at Eve. As a result, the received signal power at Eve is significantly suppressed while maintaining a distinguishable channel-domain constellation at Bob.

The detailed design of the \gls{DU-FASCC} and \gls{DC-FASCC} schemes is presented in the following subsections.

\subsection{DU-FASCC (Deliberate Uncertainty)}
\label{subsec:MRFASCC}


The proposed \gls{DU-FASCC} scheme exploits the large number of available \gls{FAS} ports to construct multiple feasible transmission realizations for each message. Unlike conventional modulation schemes, where each symbol is mapped to a single signal point, \gls{DU-FASCC} associates each message $m$ with a set of $K_{DU} \ll N$ candidate port configurations. This set is designed such that all realizations yield approximately the same received constellation point at Bob, while inducing uncertainty at Eve.

Importantly, the construction of these candidate realizations is performed by Bob using his channel knowledge only, without requiring any \gls{CSI} from Eve. 
Let
\begin{equation}
\mathcal{X}_m^{(B)} =
\left\{
x_{m,1}^{(B)}, x_{m,2}^{(B)}, \ldots, x_{m,K_{DU}}^{(B)}
\right\}
\label{eq:set_XBm}
\end{equation}
denote the set of feasible message signals 
associated with message $m$ at Bob. Each element corresponds to a different \gls{FAS} port configuration and phase compensation, but is designed to satisfy
\begin{equation}
x_{m,k}^{(B)} \approx x_{m,k'}^{(B)}, \quad \forall_{k\neq k'}.
\end{equation}

Consequently, all $K_{DU}$ realizations correspond to the same message, and the respective constellation points are almost indistinguishable from Bob's perspective.

To construct the constellation, the strongest available channel coefficient is first selected as a reference,
\begin{equation}
g_{\mathrm{ref}} = \max_{n} |g_n^{(B)}|,
\end{equation}
whose magnitude determines the radius of the target constellation. An L-ary \gls{PSK} constellation is then generated according to\footnote{Note that other target constellations can also be used, like 16-\gls{QAM}.}
\begin{equation}
x_{\mathrm{target}_m}^{(B)}
=
g_{\mathrm{ref}}
\exp
\left(
j\frac{2\pi(m-1)}{L}
\right),
\qquad
m=1,\ldots,L.
\end{equation}

For each target point $x_{\mathrm{target}_m}^{(B)}$, the $K_{DU}$ candidate realizations are obtained by selecting ports whose channel amplitudes best match the desired constellation radius. Since the subsequent phase compensation aligns the received signal with the target phase, the port-selection process is based solely on the amplitude mismatch
\begin{equation}
\left|
|g_n^{(B)}|
-
|x_{\mathrm{target}_m}^{(B)}|
\right|,
\end{equation}
where smaller values indicate a closer approximation to the target constellation point. Let $p_{m,k}$ denote the selected port associated with the $k$-th realization of message $m$.
\begin{equation}
w_{m,k} =
\exp\left[
j\left(\angle x_{\mathrm{target}_m}^{(B)} - \angle g_{p_{m,k}}^{(B)}\right)
\right].
\end{equation}

During operation, once a message $m$ is selected for transmission, the \gls{BS} randomly chooses one realization index $k \in \{1,\ldots,K_{DU}\}$. 
The corresponding port $p_{m,k}$ is activated, and the precomputed phase weight $w_{m,k}$ is applied.

The resulting message signal at receiver $u \in \{B,E\}$ is then expressed as
\begin{equation}
x_{m,k}^{(u)} = g_{p_{m,k}}^{(u)} s_{m,k} = g_{p_{m,k}}^{(u)} w_{m,k}.
\end{equation}
Consequently, the received signal model in \eqref{eq:y_rx} becomes
\begin{equation}
y^{(u)} = x_{m,k}^{(u)} + n^{(u)},
\end{equation}
and the \gls{ML} detection should be performed across all possible $x_{m,k}^{(u)}$.

At Bob, the phase weights are designed such that all realizations in $\mathcal{X}_m^{(B)}$ are aligned to the same constellation point, ensuring that different choices of $k$ do not affect symbol detection. In contrast, at Eve, the absence of channel-dependent phase compensation leads to distinct received points for different realizations of the same message.

As a result, Bob observes an approximately $L$-point constellation, whereas Eve observes up to $K_{DU}L$ distinct received points. The combination of random realization selection and mismatched phase compensation causes the realizations corresponding to the same message to be dispersed in Eve's signal space, forming an enlarged and irregular constellation whose geometry is not controlled by the transmitter, while preserving a compact and decodable structure at Bob. This behavior is illustrated in Fig.~\ref{fig:DU_constellation}, which shows the resulting \gls{DU-FASCC} channel-domain constellations observed at Bob and Eve. In this example, increasing $K_{DU}$ from $1$ to $4$ reduces the minimum Euclidean distance between constellation points at Eve from $0.78$ to $0.075$, while maintaining a minimum distance above $2$ at Bob. This demonstrates the ability of the proposed scheme to substantially degrade Eve's constellation geometry while preserving reliable detection at Bob, despite operating under a unit transmit-power constraint.

\subsection{DC-FASCC (Destructive Combining)}
\label{subsec:DCFSS}

The \gls{DC-FASCC} scheme considers a scenario when \gls{CSI} of both Bob and Eve is available. This additional knowledge is exploited to explicitly suppress the received signal at Eve to make reliable detection of the transmitted information extremely difficult, while preserving a structured constellation at Bob.


In contrast to \gls{DU-FASCC}, where security is achieved through uncertainty in the port-selection process, \gls{DC-FASCC} directly incorporates Eve's channel into the transmission design. Specifically, for each message index $m$, $K_{DC}\ll N$ \gls{FAS} ports are simultaneously activated, forming a transmission subset $\mathcal{P}_m=\{p_{m,1},\ldots,p_{m,K_{DC}}\}$. The corresponding message signal at a generic receiver $u \in {B,E}$ is given by
\begin{equation}
x_m^{(u)} = \frac{1}{\sqrt{K_{DC}}} \sum_{k=1}^{K_{DC}} g_{p_{m,k}}^{(u)} s_{m,k} = \frac{1}{\sqrt{K_{DC}}} \sum_{k=1}^{K_{DC}} g_{p_{m,k}}^{(u)} w_{m,k},
\label{eq:dcFASCC_rx}
\end{equation}

where $u \in {B,E}$ and $w_{m,k}$ denotes the complex transmit weight applied to the $k$-th port. The normalization factor $1/\sqrt{K_{DC}}$ ensures a constant total transmit power equal to unity, regardless of the number of active ports.

The transmit weights are chosen to align the selected channels toward the desired constellation point at Bob, while the subset-selection procedure exploits Eve's \gls{CSI} to identify port combinations that simultaneously reduce the received signal strength at Eve. 

For the first constellation point, candidate port subsets are evaluated according to the metric
\begin{equation}
\Gamma_1 =
\frac{|x_1^{(B)}|^\alpha}
{|x_1^{(E)}|+\epsilon},
\end{equation}
where $\epsilon$ is a small constant introduced for numerical stability. This metric promotes constructive combining at Bob while simultaneously ensuring that the first selected port at Eve will be close to $0$.

For the remaining constellation points, the subset-selection procedure seeks to maximize the minimum Euclidean distance between constellation points at Bob while minimizing the corresponding separation at Eve. Specifically, candidate subsets are ranked according to
\begin{equation}
\Gamma_m = \frac{d_B^\alpha} {d_E+\epsilon}, \qquad m=2,\ldots,L,
\end{equation}
where
\begin{equation}
d_B= \min_{i<m} |x_m^{(B)}-x_i^{(B)}|
\end{equation}
and
\begin{equation}
d_E= \min_{i<m} |x_m^{(E)}-x_i^{(E)}|.
\end{equation}

The parameter $\alpha \geq 0$ controls the tradeoff between reliability and secrecy. Larger values prioritize constellation separation at Bob, whereas smaller values place greater emphasis on reducing the distinguishability of the received signals at Eve.

\begin{figure}[h]
    \centering

    \begin{subfigure}{\linewidth}
        \centering
        \includegraphics[width= \linewidth,trim=4cm 0cm 4cm 0cm,clip]{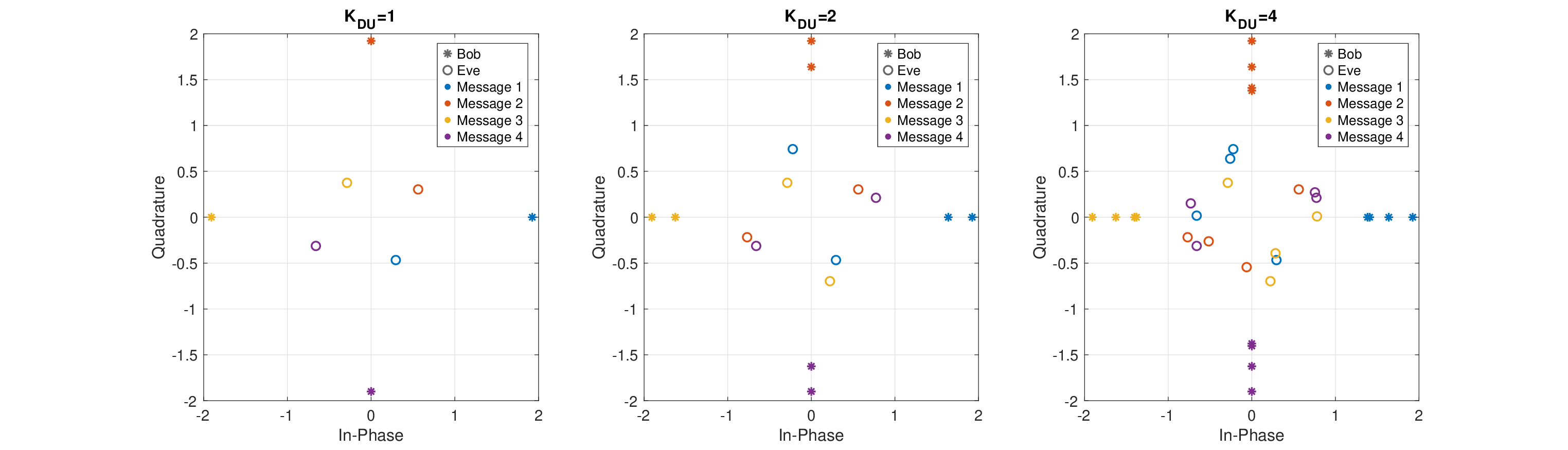}
        \caption{DU}
        \label{fig:DU_constellation}
    \end{subfigure}

    \vspace{0.5em}

    \begin{subfigure}{\linewidth}
        \centering
        \includegraphics[width= \linewidth,trim=4cm 0cm 4cm 0cm,clip]{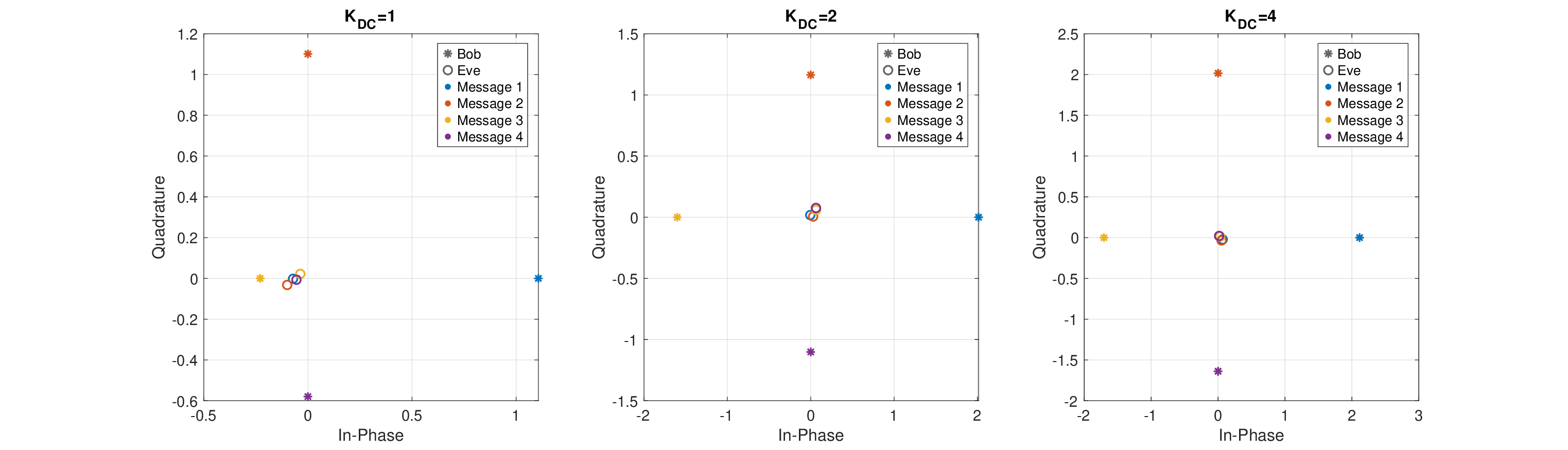}
        \caption{DC ($\alpha=1$)}
        \label{fig:DC_constellation}
    \end{subfigure}
    \caption{Effective received constellation $x^{(u)}_m$ for Bob and Eve with different values of $K_{DU}$ and $K_{DC}$.}
    \label{fig:constellation_schemes}
\end{figure}

Fig.~\ref{fig:DC_constellation} illustrates the resulting \gls{DC-FASCC} constellation structure for different values of $K_{DC}$ using $\alpha=1$. It can be observed that the received power at Eve remains below $0.005$ in all considered cases, confirming the effectiveness of the proposed destructive-combining strategy in suppressing the eavesdropper's signal. Although increasing $K_{DC}$ provides a modest additional reduction in Eve's received power, its primary benefit lies in the constructive combining gain achieved at Bob. In the illustrated example, the average received power at Bob increases from $0.73$ for $K_{DC}=1$ to $3.7$ for $K_{DC}=4$, resulting in a substantially more distinguishable constellation while maintaining a highly attenuated signal at Eve.


\subsection{Design Considerations}
\label{subsec:design_considerations}

In addition to the scheme-specific design procedures described previously, several common considerations apply to both proposed \gls{FASCC} schemes.

Both proposed \gls{FASCC} schemes rely on the selection of multiple \gls{FAS} ports for each transmitted message. To ensure that these ports provide meaningful spatial diversity, the selected port indices are constrained by a minimum spacing ($d_{\min}$). Since the spatial correlation between \gls{FAS} ports decreases with their separation, this constraint prevents the selection of highly correlated ports that would otherwise produce nearly identical channel realizations. Consequently, the selected ports exhibit sufficiently distinct channel responses, which increases the variability of the received signal observed by Eve and improves the effectiveness of the proposed secure transmission strategies.

In both algorithms, the port-selection and weight-design procedure uses a greedy search over random candidate subsets, guided by a secrecy-aware metric that favors large constellation separation at Bob and reduced distinguishability at Eve. Although this approach does not guarantee global optimality, it provides a computationally efficient approximation that enables practical implementation even for large numbers of available \gls{FAS} ports.

In addition, both schemes can be combined with artificial transmit noise to further degrade Eve's reception. Specifically, a zero-mean complex Gaussian perturbation
$z \sim \mathcal{CN}(0,\sigma_z^2)$
is independently generated for each transmission and is unknown to both Bob and Eve. The perturbation is added to the transmit weights, such that the transmitted signal becomes
\begin{equation}
    s_{m,k}=w_{m,k}+z.
\end{equation}

The artificial perturbation is controlled by a single parameter, namely its variance $\sigma_z^2$, and does not require knowledge of Eve's channel. Since the proposed \gls{FASCC} schemes are designed to provide either a favorable constellation geometry (\gls{DU-FASCC}) or a received-power advantage (\gls{DC-FASCC}) at Bob, moderate levels of transmit noise can typically be tolerated with limited performance degradation. In contrast, Eve already experiences either increased constellation ambiguity or signal suppression, making her reception significantly more sensitive to the additional perturbation. Consequently, increasing $\sigma_z^2$ provides an additional security--reliability tradeoff. Moreover, because the perturbation is introduced at the transmitter, its impact cannot be eliminated solely through a higher received \gls{SNR}, allowing the proposed approach to remain effective even when Eve is located closer to the transmitter than Bob. 

Finally, Eve is assumed to know the modulation alphabet and the bit-to-symbol mapping adopted by the legitimate system, consistent with Kerckhoffs' principle\cite{Menezes1996}. Therefore, the reported results do not rely on security through obscurity. In practice, however, identifying the underlying transmission structure and the associated message realizations may require additional observation and estimation effort, particularly for the proposed \gls{DU-FASCC} scheme, where multiple channel realizations correspond to the same message.

\subsection{Performance Metrics}
\label{subsec:metrics}


The performance of the proposed schemes is evaluated in terms of the \gls{BER}, secrecy gap, received signal power, and mutual information.

The \gls{BER} is adopted as the primary reliability metric and is evaluated separately for Bob and Eve. Since the proposed designs aim to maximize the distinguishability of the channel-domain constellation at Bob while reducing it at Eve, the BER directly quantifies both communication reliability and security.

The secrecy gap quantifies the difference between the \gls{SNR} required by Bob to achieve reliable communication and the \gls{SNR} required by Eve to successfully decode the transmitted information. Let $\beta_B$ and $\beta_E$ denote the target \gls{BER}s for Bob and Eve, respectively. The secrecy gap is defined as
\begin{equation}
G_s = \gamma_E(\beta_E) - \gamma_B(\beta_B),
\end{equation}
where $\gamma_B(\beta_B)$ is the minimum \gls{SNR} at which Bob achieves a \gls{BER} below $\beta_B$, and $\gamma_E(\beta_E)$ is the maximum \gls{SNR} for which Eve's \gls{BER} remains above $\beta_E$. Larger secrecy gaps indicate a wider operating region in which reliable communication is maintained at Bob while preventing successful decoding at Eve.

The average received signal power is given by
\begin{equation}
P_{x^{(u)}} = \mathbb{E}
\left[ |x_m^{(u)}|^2 \right],
\end{equation}
where the expectation is taken over all transmitted messages and channel realizations. This metric is particularly relevant for assessing the effectiveness of the proposed suppression mechanisms. A low received power at Eve reduces the probability of detecting the existence of a communication link, while a high received power at Bob contributes to reliable message recovery.

Finally, the secrecy performance is evaluated through the mutual information secrecy rate. Let $I_B$ and $I_E$ denote the mutual information between the transmitted message and the received observations at Bob and Eve, respectively.

Let $M$ denote the transmitted message index. Assuming equiprobable messages, i.e., $p(m)=1/L$, the mutual information is given by
\begin{equation}
I(M;Y) = \sum_{m=1}^{L} \frac{1}{L} \int p(y|m) \log_2 \left( \frac{p(y|m)} {p(y)} \right) dy,
\end{equation}
where
\begin{equation}
p(y) = \sum_{m=1}^{L} \frac{1}{L}p(y|m).
\end{equation}

For the \gls{DU-FASCC} scheme, the conditional distribution is given by
\begin{equation}
p(y|m) = \frac{1}{K_{DC}} \sum_{k=1}^{K_{DC}} p(y|x_{m,k}),
\end{equation}
whereas for \gls{DC-FASCC} each message is associated with a single realization and therefore $p(y|m) = p(y|x_m)$.

The conditional distribution $p(y|m)$ incorporates both the receiver noise and the artificial transmit perturbation introduced by the considered transmission scheme.

The mutual information secrecy rate is then defined as
\begin{equation}
R_s = \left[ I_B-I_E \right]^+,
\end{equation}
where $[x]^+=\max(x,0)$.

\section{Performance Results}
\label{sec:Performance}

The simulations consider a rich-scattering environment and, unless otherwise stated, a one-dimensional \gls{FAS} deployed at the \gls{BS} with aperture $W=10$ and $N=200$ uniformly spaced ports. The \gls{BS} conveys $2$ bits per transmission, corresponding to an $L=4$ channel-domain constellation. Bob and Eve are equipped with a single antenna and experience independent fading channels. For all considered schemes and parameter settings, the total transmit power is normalized to unity, ensuring a fair comparison of the resulting constellation geometry and secrecy performance.


Although $K_{DU}$ and $K_{DC}$ have different physical interpretations, representing the number of candidate realizations per message in \gls{DU-FASCC} and the number of simultaneously active ports in \gls{DC-FASCC}, respectively, they both control the amount of spatial diversity exploited by the transmission design. Therefore, for simplicity, we use the common notation $K$ when presenting simulation results and discussing general performance trends.

The following transmission schemes are evaluated for different values of $K$:
\begin{itemize}
\item \textbf{\gls{DU-FASCC}}: Each message is associated with $K$ candidate transmission realizations, and one realization is randomly selected during transmission. The scheme requires only Bob's CSI and increases security by creating additional uncertainty at Eve.

\item \textbf{\gls{DC-FASCC}}: For each message, $K$ ports are simultaneously activated and jointly optimized using Bob's and Eve's CSI. The parameter $\alpha=1$ is adopted throughout the simulations, providing a balanced tradeoff between maximizing constellation distinguishability at Bob and suppressing the received signal at Eve.
\end{itemize}

\begin{figure}[h]
    \centering
    \includegraphics[width=1.0\linewidth]{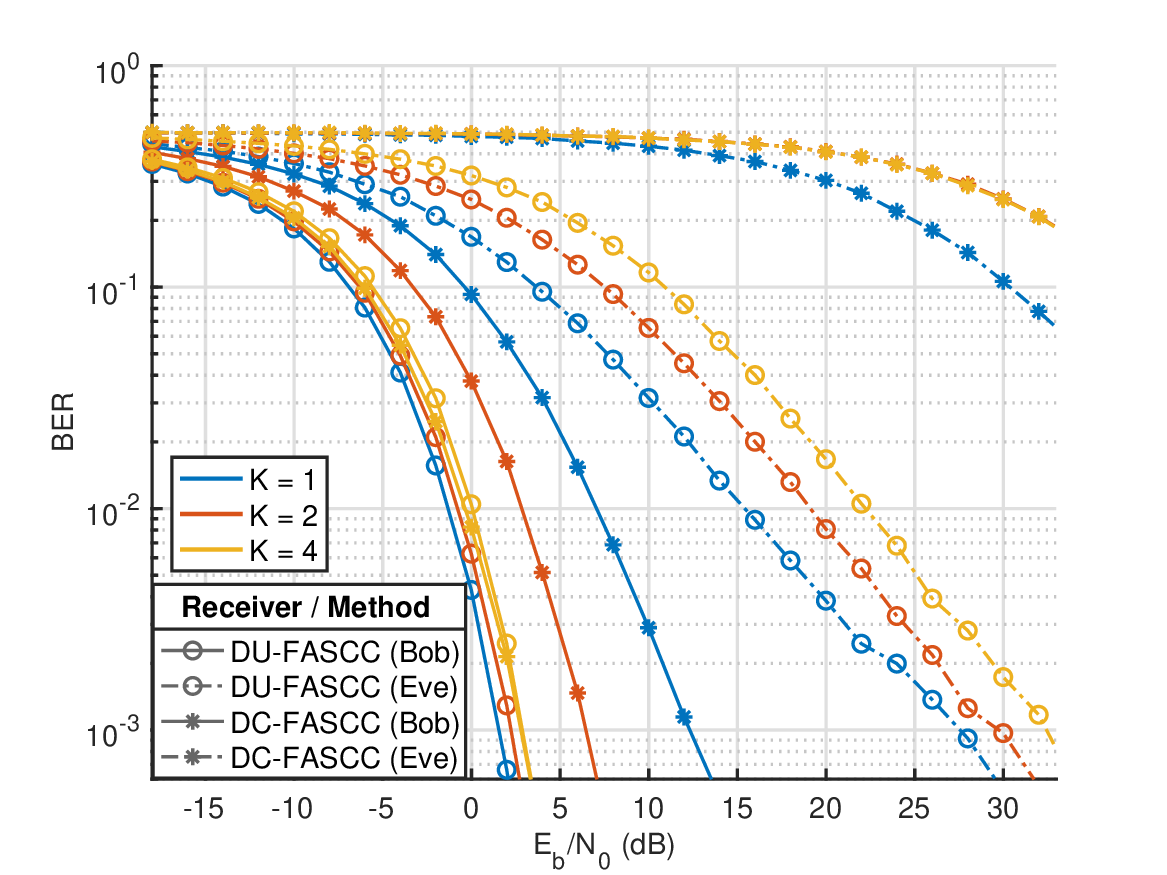}
    \caption{BER performance of Bob and Eve for different $K$ values without artificial transmitted noise.}
    \label{fig:BER}
\end{figure}

We begin by considering the case without artificial transmit noise, i.e., $\sigma_z^2=0$. Fig.~\ref{fig:BER} shows the \gls{BER} as a function of $E_b/N_0$ for different values of $K$. As expected, \gls{DU-FASCC} provides better \gls{BER} performance at Bob since it is designed solely to maximize constellation distinguishability at the legitimate receiver. In contrast, \gls{DC-FASCC} sacrifices part of this gain to suppress Eve's received signal, resulting in a higher \gls{BER} at Bob but significantly degraded performance at Eve. Furthermore, increasing $K$ improves security for both schemes. In \gls{DU-FASCC}, larger values of $K$ increase the ambiguity of Eve's observations by creating additional transmission realizations for each message, whereas in \gls{DC-FASCC} they provide additional degrees of freedom for destructive combining at Eve.

\begin{figure}[h]
    \centering
    \includegraphics[width=1.0\linewidth]{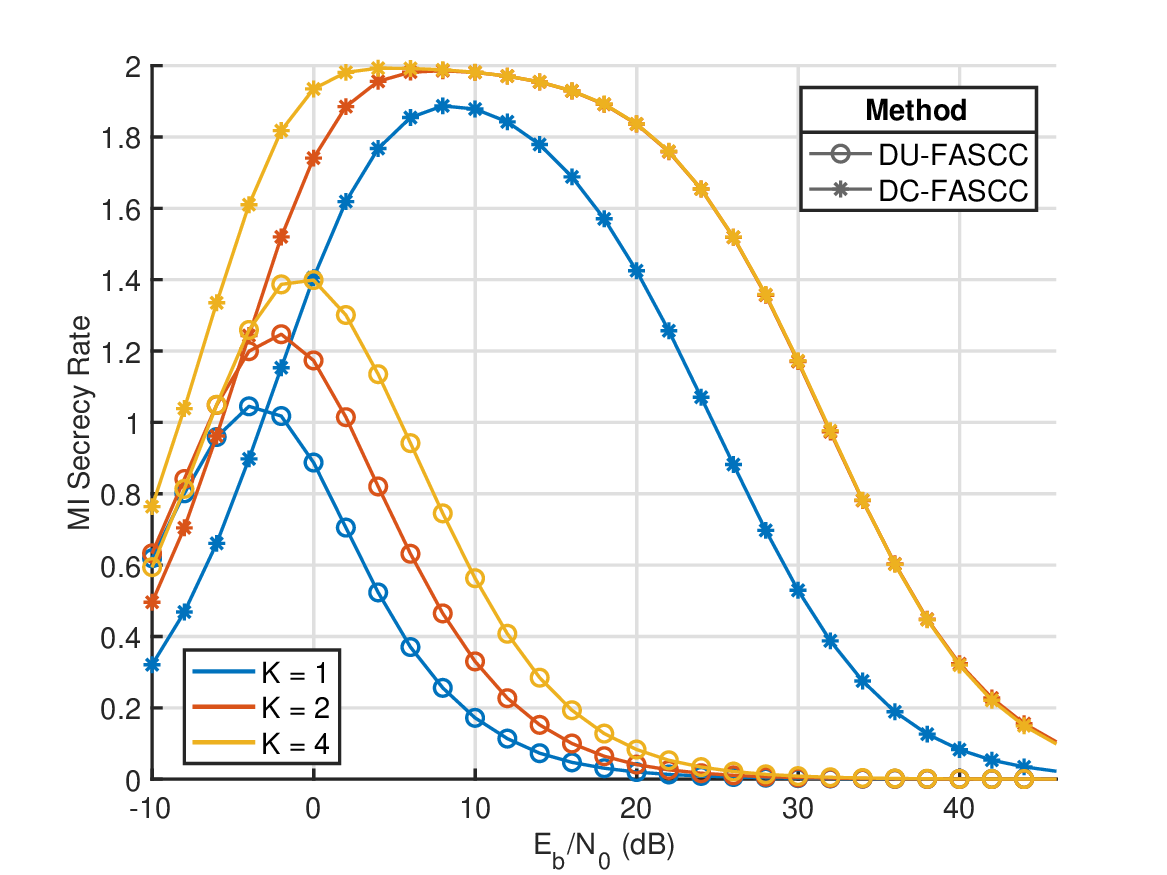}
    \caption{Mutual information secrecy rate performance for different $K$ values without artificial transmitted noise.}
    \label{fig:MI}
\end{figure}

The security benefits of the proposed schemes are more clearly illustrated in Fig.~\ref{fig:MI}, which shows the mutual-information secrecy rate defined in Section~\ref{subsec:metrics} as a function of $E_b/N_0$ for different values of $K$. This metric provides a conservative assessment of secrecy performance, as it assumes optimal decoding capabilities at Eve and quantifies the difference between the information rates achievable at Bob and Eve.

It can be observed that \gls{DC-FASCC} achieves substantially higher secrecy rates than \gls{DU-FASCC}. In particular, \gls{DC-FASCC} maintains secrecy rates close to the maximum achievable value of $2$ bits/symbol over a wide \gls{SNR} range, whereas \gls{DU-FASCC} reaches a peak secrecy rate of approximately $1.4$ bits/symbol. This behavior highlights the benefit of exploiting Eve's \gls{CSI} to actively suppress the received signal at the eavesdropper. For both schemes, increasing $K$ improves the secrecy rate; however, as the \gls{SNR} becomes very large, the secrecy rate gradually decreases. In the absence of artificial transmit noise, both Bob and Eve eventually operate in a nearly noise-free regime, allowing Eve to increasingly resolve the transmitted messages despite the proposed security mechanisms.

The results in Fig.~\ref{fig:BER} and Fig.~\ref{fig:MI} reveal an important limitation of the noise-free case. Although both \gls{FASCC} schemes provide substantial secrecy gains at moderate \gls{SNR}, the secrecy rate eventually decreases at high \gls{SNR} because Eve can reliably distinguish the received constellation points when addictive noise becomes negligible.

\begin{figure}[h]
    \centering
    \includegraphics[width=1.0\linewidth]{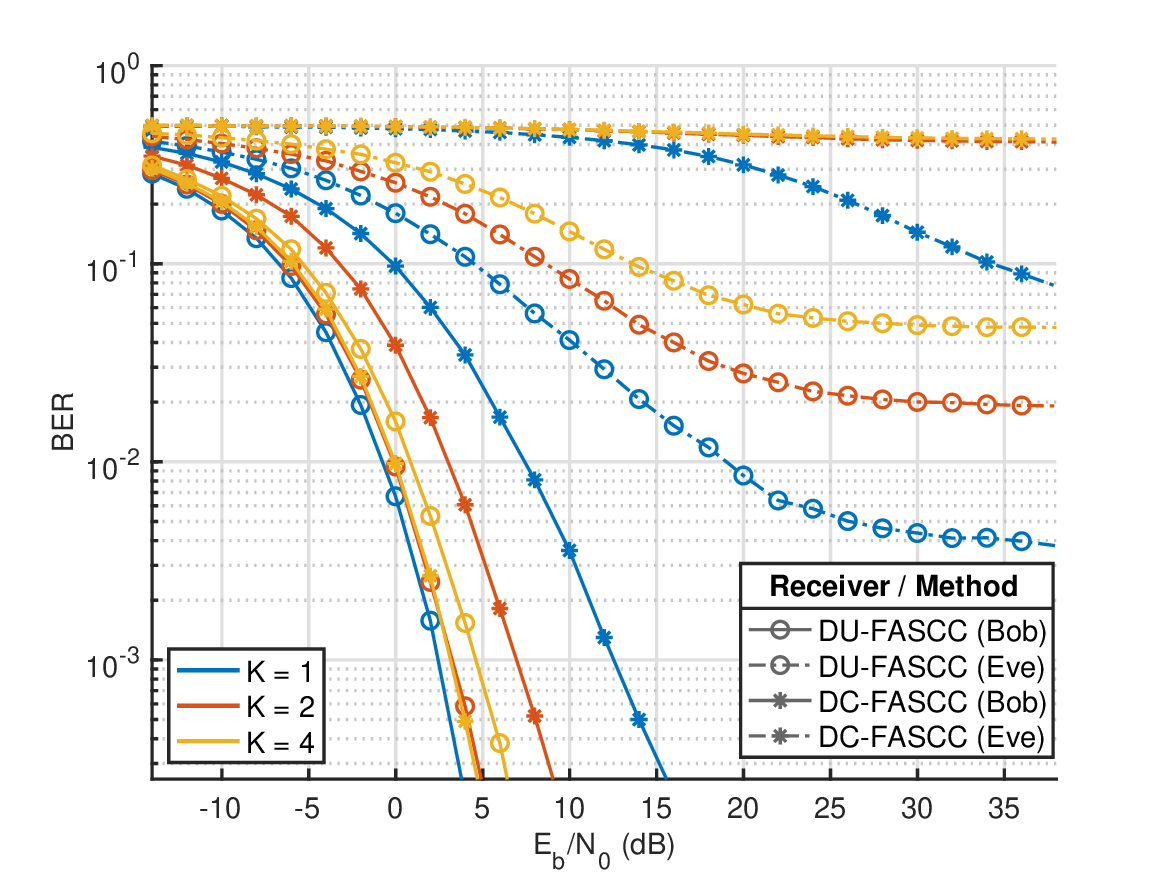}
    \caption{BER performance of Bob and Eve for different $K$ values with $\sigma^2_z=0.15$ for \gls{DU-FASCC} and \gls{DC-FASCC} schemes.}
    \label{fig:BER_txnoise}
\end{figure}

\begin{figure}[h]
    \centering
    \includegraphics[width=1.0\linewidth]{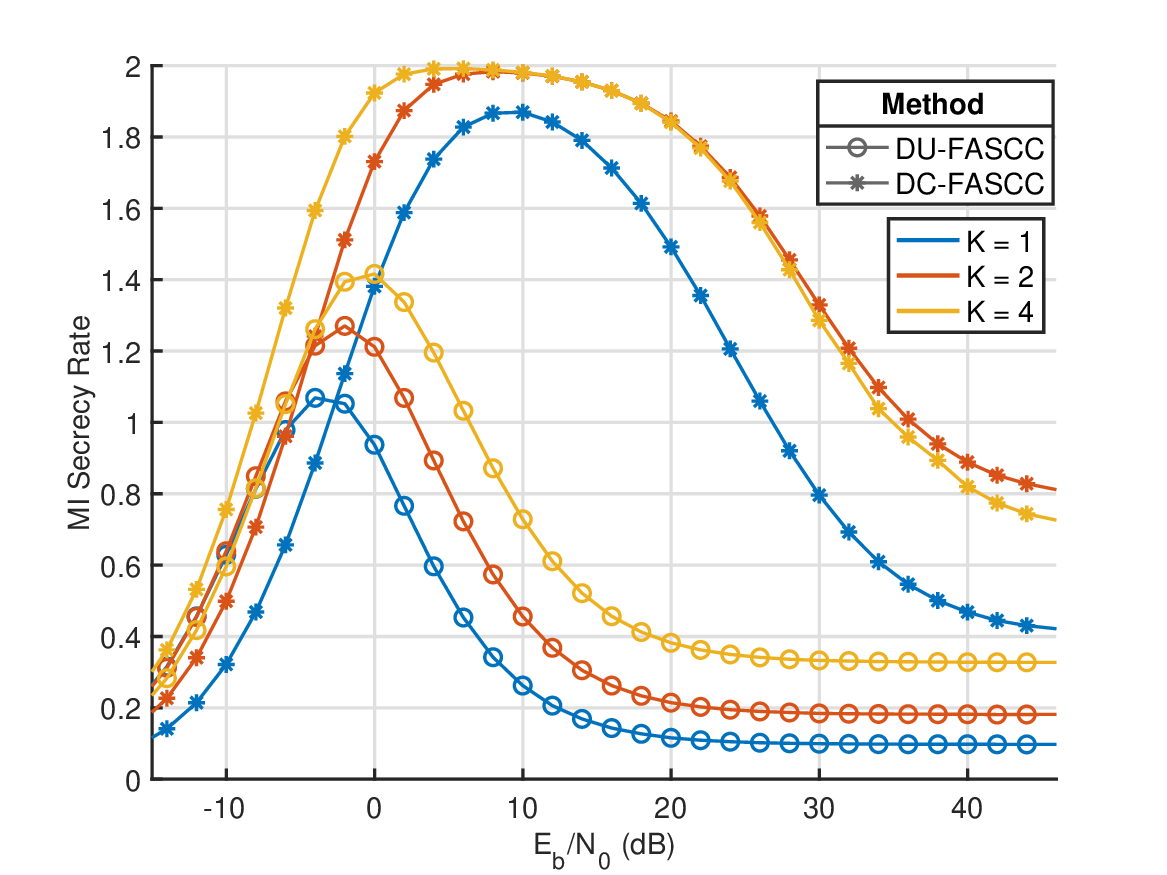}
    \caption{Mutual information secrecy rate performance for different $K$ values with $\sigma^2_z=0.15$ for \gls{DU-FASCC} and \gls{DC-FASCC} schemes.}
    \label{fig:MI_txnoise}
\end{figure}

To address this issue, Fig.~\ref{fig:BER_txnoise} and Fig.~\ref{fig:MI_txnoise} repeat the previous experiments with artificial transmit noise of variance $\sigma_z^2=0.15$. The impact is immediately visible in Fig.~\ref{fig:BER_txnoise}. While Bob experiences only a modest performance degradation, Eve's \gls{BER} exhibits a clear error floor that cannot be eliminated by increasing the received \gls{SNR}. In particular, the error floor is approximately $0.07$ for \gls{DU-FASCC}, whereas \gls{DC-FASCC} maintains an almost random detection performance with a \gls{BER} close to the optimal value of $0.5$. It is worth noting that these gains come at the cost of increased complexity, as \gls{DC-FASCC} requires Eve's \gls{CSI} and multiple simultaneously active ports, whereas \gls{DU-FASCC} relies only on Bob's \gls{CSI} and a single active port per transmission.

This behavior is further reflected in Fig.~\ref{fig:MI_txnoise}. Unlike the noise-free case, where the secrecy rate approaches zero at high \gls{SNR}, the artificial transmit perturbation preserves a non-zero secrecy rate even when Eve is located significantly closer to the \gls{BS} than Bob. For the considered setup, the secrecy rate remains close to $0.4$ for \gls{DU-FASCC} and $0.8$ for \gls{DC-FASCC}. These results demonstrate that the proposed perturbation creates uncertainty that cannot be eliminated through increased received power alone, allowing both \gls{FASCC} schemes to maintain meaningful secrecy even when Eve enjoys a substantial \gls{SNR} advantage.

\begin{figure}[h]
    \centering
    \includegraphics[width=1.0\linewidth]{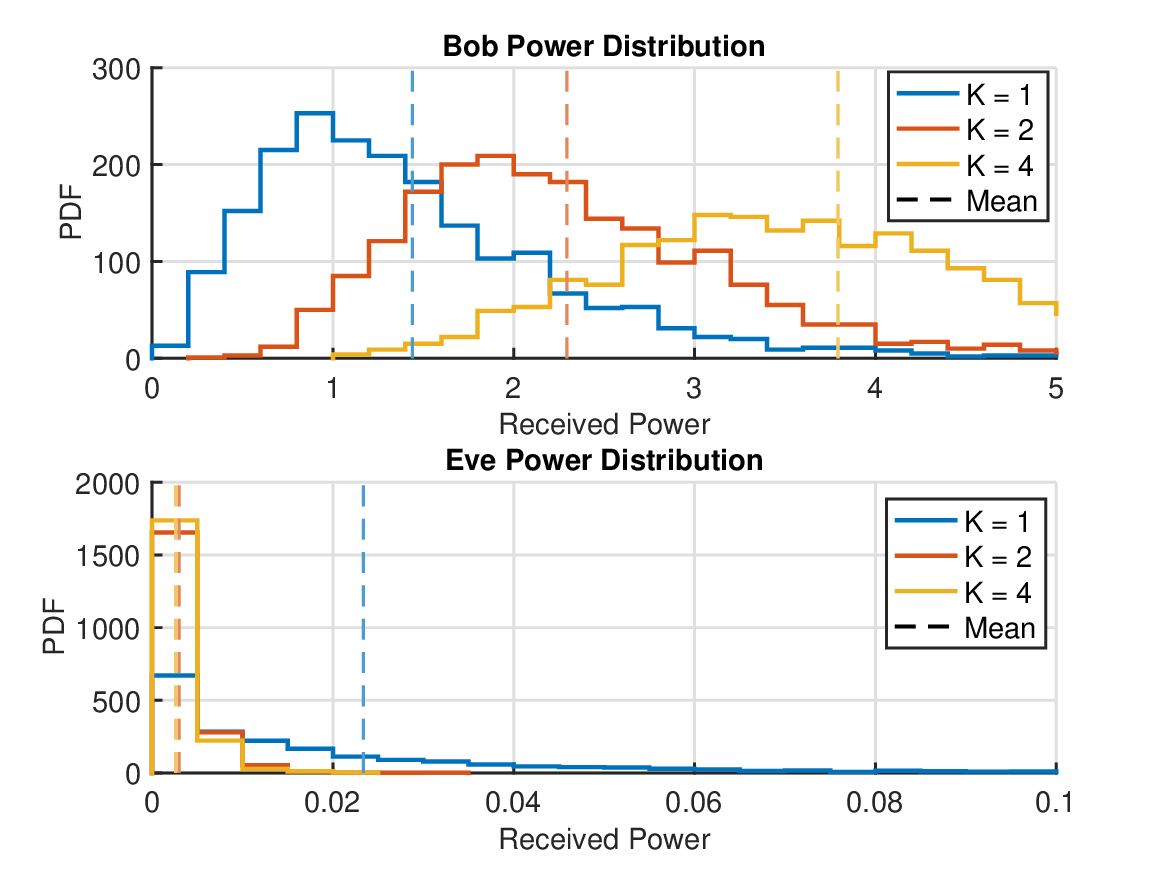}
    \caption{Received power distribution for different $K$ values for \gls{DC-FASCC}.}
    \label{fig:Rxpower}
\end{figure}

Fig.~\ref{fig:Rxpower} shows the distribution of the received power at Bob and Eve for \gls{DC-FASCC} under different values of $K$. Recall that the total transmit power is normalized to unity. As expected, increasing $K$ provides additional degrees of freedom for the port-selection process, resulting in a simultaneous increase in Bob's received power and a reduction in Eve's received power. This trend is consistent with the constellation behavior observed in Fig.~\ref{fig:DC_constellation}, where larger values of $K$ produced more distinguishable constellation points at Bob while further suppressing the received signal at Eve. Overall, these results confirm the effectiveness of the proposed \gls{DC-FASCC} strategy in exploiting constructive combining at Bob and destructive combining at Eve, significantly reducing the eavesdropper's ability to detect the presence of a transmission and recover the transmitted information.

\begin{figure}[h]
    \centering
    \includegraphics[width=1.0\linewidth]{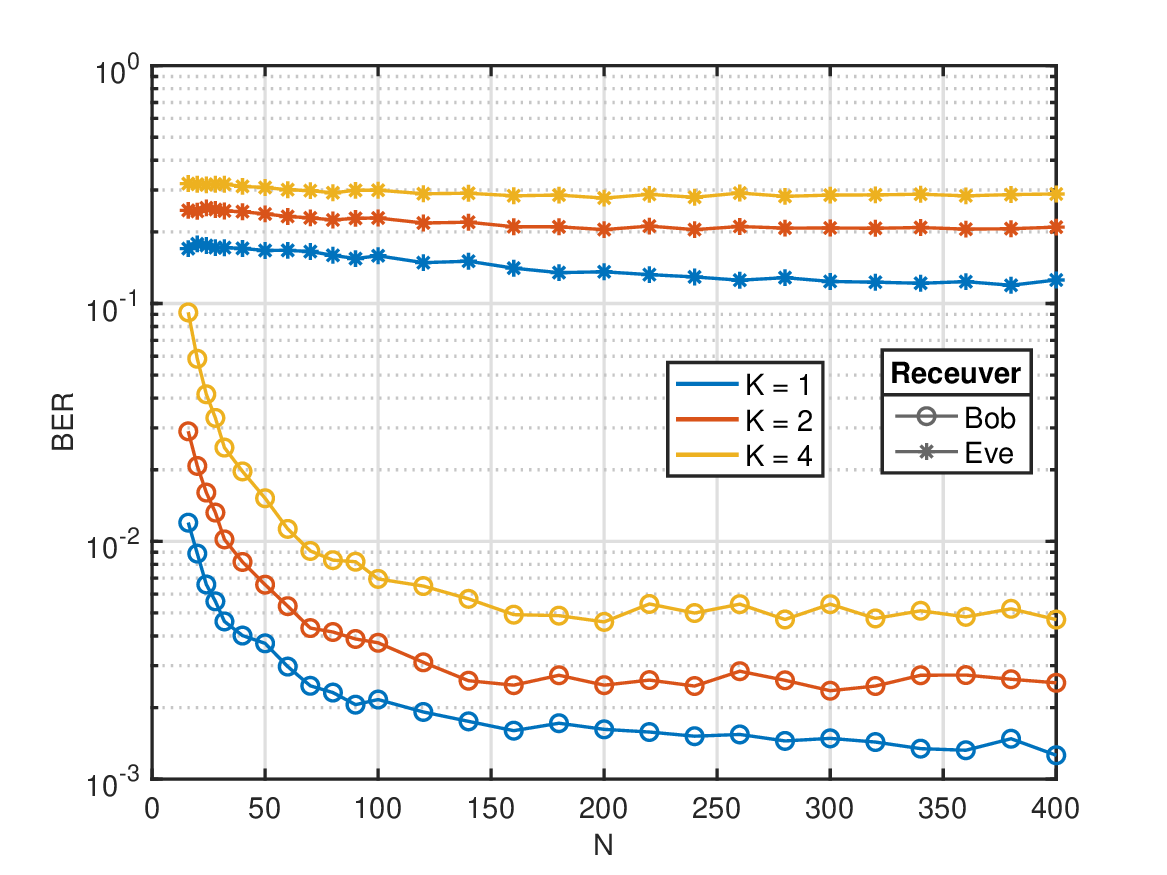}
    \caption{BER as a function of the number of \gls{FAS} ports $N$, for different $K$ values with $\sigma^2_z=0.15$ for \gls{DU-FASCC}.}
    \label{fig:BER_N}
\end{figure}

Finally, we analyze the impact of the \gls{FAS} architecture on system performance. Fig.~\ref{fig:BER_N} presents the \gls{BER} at $\mathrm{SNR}=5$dB as a function of the number of available \gls{FAS} ports $N$. We focus on the \gls{DU-FASCC} scheme with artificial transmit noise variance $\sigma_z^2=0.15$. As expected, increasing $N$ improves Bob's performance by providing a larger set of candidate channel realizations for port selection. In particular, the performance continues to improve well beyond the spatial Nyquist sampling rate ($N \approx 21$ for the considered aperture), which would correspond to a conventional fixed antenna array with approximately half-wavelength antenna spacing. This behavior highlights a key advantage of the proposed \gls{FAS}-based framework: unlike conventional antenna arrays, whose spatial degrees of freedom are largely determined by the aperture size and not the number of ports, in our system increasing the number of available \gls{FAS} ports provides additional channel diversity that can be exploited to construct more favorable channel-domain constellations. As a result, significant gains are observed even when the aperture remains fixed. In contrast, Eve's \gls{BER} remains nearly unchanged with increasing $N$, since the additional channel realizations are selected using Bob's channel information and therefore primarily benefit the legitimate receiver.

The figure also shows that increasing $K$ degrades the \gls{BER} performance of both Bob and Eve. This behavior is expected in \gls{DU-FASCC}, as larger values of $K$ introduce additional transmission uncertainty by associating more channel realizations with each message. 

\begin{figure}[h]
    \centering
    \includegraphics[width=1.0\linewidth]{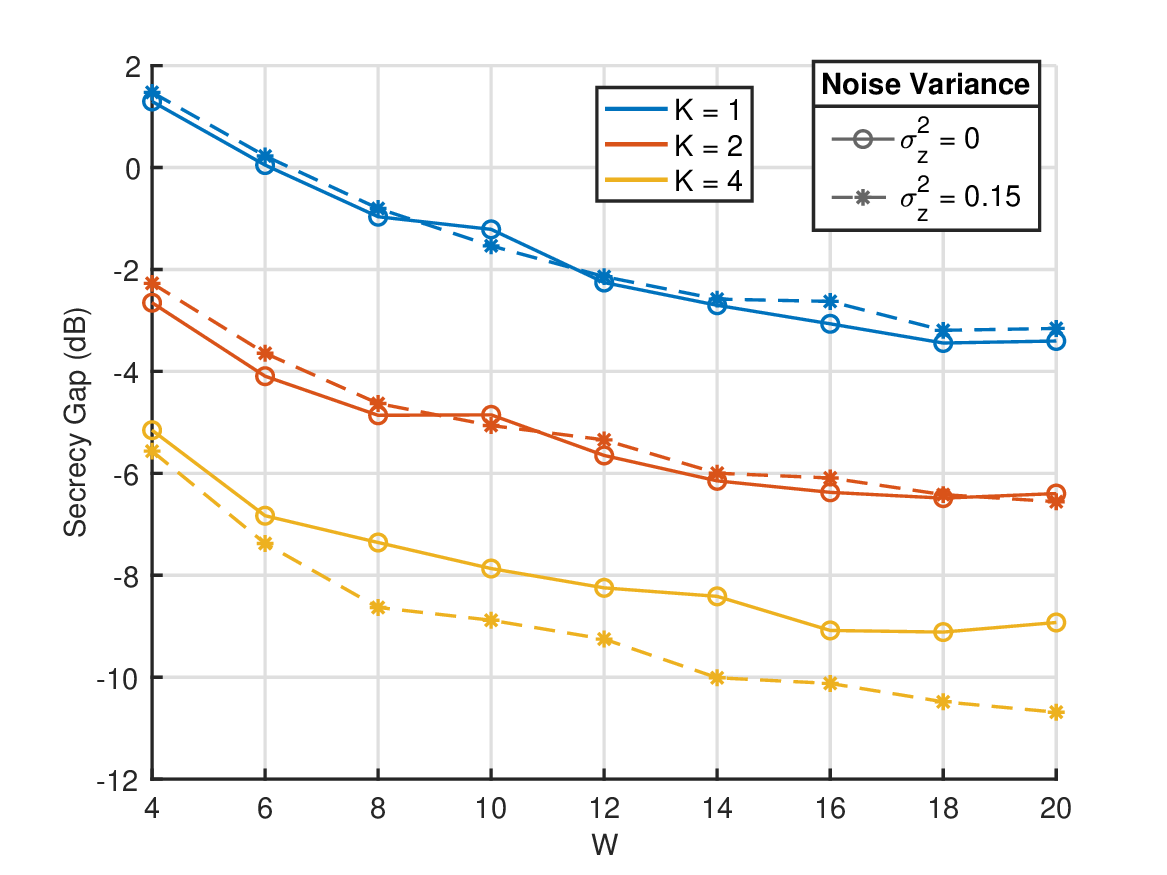}
    \caption{Secrecy gap as a function of the antenna size $W$ for different $K$ values with $\sigma^2_z=\{0,0.15\}$ for \gls{DU-FASCC}.}
    \label{fig:secrecy_gap}
\end{figure}

Having analyzed the impact of the number of available \gls{FAS} ports $N$, we now investigate the effect of the aperture size $W$. Fig.~\ref{fig:secrecy_gap} evaluates the secrecy gap, defined as the difference in \gls{SNR} required for Bob to achieve a target \gls{BER} of $\beta_B=10^{-3}$ and for Eve to achieve a target \gls{BER} of $\beta_E=10^{-1}$, as a function of $W$ for different values of $K$. The analysis focuses on \gls{DU-FASCC}, since Fig.~\ref{fig:BER_txnoise} shows that Eve is unable to reach the target \gls{BER} under \gls{DC-FASCC}, resulting in an effectively unbounded secrecy gap. As expected, increasing $W$ improves the secrecy gap by providing a larger set of channel realizations and greater spatial diversity, which enhances the port-selection process and the resulting constellation geometry. Similarly, larger values of $K$ consistently increase the secrecy gap by introducing additional ambiguity into Eve's observations. Finally, the benefits of artificial transmit noise become more pronounced for larger $K$, where the inherent robustness of the proposed design allows the additional perturbation to further degrade Eve's reception while having only a limited impact on Bob's detection performance.

\section{Conclusions}
\label{sec:conclusions}

This paper proposed a \gls{FASCC} framework for downlink secure communications, where information is conveyed through channel-domain constellations generated via \gls{FAS} port selection and phase adaptation. 


Two schemes were developed: \gls{DU-FASCC}, which relies only on Bob's \gls{CSI} and increases Eve's uncertainty through multiple possible transmission realizations, and \gls{DC-FASCC}, which additionally exploits Eve's \gls{CSI} to suppress its received signal via destructive combining with the goal of covert communication. Numerical results show significant improvements in \gls{BER} and mutual-information secrecy performance, being even able to maintain secrecy at high \gls{SNR}.

\bibliographystyle{IEEEtran}
\bibliography{rw}

\end{document}